\documentclass[epj]{svjour}
\RequirePackage[utf8]{inputenc}
\RequirePackage{xspace}
\RequirePackage{graphicx}
\RequirePackage{amsmath}
\RequirePackage{amssymb}
 \RequirePackage{xcolor}
 \RequirePackage[sort&compress]{natbib}
 \RequirePackage{hepunits}
 \RequirePackage{enumitem}
 \usepackage{tikz}
 \usepackage[compat=1.1.0]{tikz-feynman}
\usepackage[colorlinks=true,linkcolor=blue]{hyperref}%
 \usepackage{definitions}%
 \usepackage{comments}%
 \usepackage{mathtools}
 \usepackage{slashed}
 \usepackage{mathrsfs}
 \usepackage{changes}
 \usepackage{orcidlink}
\begin{document}
\title{Tackling the DVCS inverse problem with finite element method}
\author{Marija Čuić\inst{1}\inst{2} \and Cédric Mezrag\inst{1}\inst{2} \and José Manuel Morgado Chávez\inst{3}
}                     
%
%
\institute{Université Paris-Saclay - CEA - IRFU, F-91191, Gif-sur-Yvette, France \and AIDAS, CEA, F-91191, Gif-sur-Yvette, France \and Department of Theoretical Physics and IFIC, University of Valencia and CSIC,
E-46100, Valencia, Spain}
\date{Received: date / Revised version: date}
%
\abstract{
  In this paper, we tackle the problem of shadow GPDs, arising from the non-invertibility of the measured $2\to 2$ exclusive processes at large virtualities (DVCS, TCS and DVMP).
  This problem arose five years ago as the main source of theoretical uncertainties, potentially dominating experimental ones by orders of magnitude.
  We show here that the deconvolution ambiguity can be elegantly eliminated by relying on finite element method.
  The problem is essentially converted at any finite discretisation area $a$ from a non-invertible to an ill-conditioned one.
  Given experimental uncertainties, this yields a theoretical lower bound on the discretisation area $a_\textrm{min}$, that can be interpreted as the maximal theoretical resolution one can obtain on the reconstruction of the lightcone momentum fractions.
  Associated with the local polynomial degree $p$, the pair $(a,p)$ provides an insightful and easily understandable regularisation framework, allowing one to systematically improve the description of the momentum fractions dependence of GPDs with the pQCD order at which the hard kernels of exclusive processes are computed.
} 
\maketitle
\section{Introduction}
\label{intro}

Generalised Parton Distributions were introduced in the 1990s independently in Refs. \cite{Mueller:1998fv,Ji:1996ek,Ji:1996nm,Radyushkin:1996ru,Radyushkin:1997ki}, and quickly raised interest for three main reasons: they encode the 3D density maps of quarks and gluons within hadrons \cite{Burkardt:2000za,Diehl:2002he}, they are connected to the energy-momentum tensor (EMT) \cite{Ji:1996ek,Polyakov:2002yz} and most importantly, they are experimentally accessible in deep exclusive reactions such as deeply virtual Compton scattering (DVCS) \cite{Ji:1996nm,Ji:1998xh,Collins:1998be}.
However, this connection is not perfect, in the sense that it has been shown in Ref. \cite{Bertone:2021yyz} that the connection between GPDs and DVCS experimental data is not invertible (note the early discussions in that respect held in Refs. \cite{Freund:1999xf,Diehl:2003ny,Kumericki:2008di}).
Indeed, the DVCS amplitude is obtained through the convolution of a coefficient function, computed in perturbative QCD, with GPDs.
This convolution has a non-trivial kernel, spanning a set of candidate functions that cannot be probed experimentally.
They were coined ``shadow GPDs'' in Ref. \cite{Bertone:2021yyz}.
As a result, at a time when experimental facilities such as JLab 12 are starting to deliver precise DVCS data, theoretical uncertainties generated by the deconvolution problem may become dominant (see for instance the impact spanning several orders of magnitude in \cite{Bertone:2021yyz,Moffat:2023svr}), spoiling our understanding of the hadron multidimensional structure.

In this paper, we will take advantage of the efforts performed during the last decade regarding the finite element description of generalised parton distributions \cite{Chouika:2017dhe,Chouika:2017rzs,Chavez:2021llq,DallOlio:2024vjv}, originally developed to tackle the continuation from the Dokshitzer-Gribov-Lipatov-Altarelli-Parisi (DGLAP) to the Efremov-Radyushkin-Brodsky-Lepage (ERBL) kinematic regions in models based on the overlap of Lightfront Wave Functions (LFWFs).
This discretisation effort reveals itself in fact a very elegant way to mitigate the impact of shadow GPDs.
Since these are polynomial functions, they are approximated by any finite element technique.
The discretisation error $\delta f_{\textrm{shadow}}$ has no reason to belong to the kernel of the coefficient function, making the discretised version of shadow GPDs visible in experimental data.
With this idea at hand, the discretisation parameter, which here is our elementary area $a$, has to reach a compromise between two opposite requirements:
\begin{enumerate}
\item In order to approximate the GPD as well as possible, one would like to have $a$ as small as possible;
\item In order to be able to clearly distinguish shadow GPDs in experimental data, one would like to have $a$ as large as possible. 
\end{enumerate}
The aim of this paper is to investigate the trade-off between these two opposite requirements, yielding a tentative definition of $a_{\textrm{min}}$, the minimal area at which one can still be sensitive to shadow GPDs being given the experimental precision of the data.

To reach this target, our paper is organised as follows. In Sec. \ref{sec:Def}, we recall the theoretical definitions and tools that will be used throughout the paper. In Sec. \ref{sec:Discretise}, we discuss the discretisation of the double distribution and benchmark our new algorithms. In Sec. \ref{sec:Deconvolution}, we tackle the deconvolution problem and discuss in detail the trade-off between sensitivity to shadow GPDs and convergence toward the targeted function. Finally we provide the reader with our conclusions.


\section{Definition and properties of GPDs}
\label{sec:Def}
\subsection{Definition of GPDs and connection with DVCS amplitude}

Following Ref. \cite{Diehl:2003ny}, GPDs are defined in terms of non-local matrix elements. In the case of nucleon chiral-even and unpolarised GPDs, one has:
\begin{align}
  & \frac{1}{2}\int \frac{\textrm{d}z^-}{2\pi} e^{ixP^+z-} \bra{p_2}\bar{\psi}^q\left(-\frac{z^-}{2}\right)\gamma^+ \psi^q\left(\frac{z^-}{2}\right) \ket{p_1}\nonumber \\
  = & \frac{1}{2P^+}\bigg[\bar{u}(p_2)\gamma^+u(p_1) H^q(x,\xi,t) \nonumber \\
  & \left. \qquad \qquad + \bar{u}(p_2) \frac{i\sigma^{+\alpha}\Delta_\alpha}{2M}u(p_1)E^q(x,\xi,t) \right],
\end{align}
where $P = \frac{p_1+p_2}{2}$, $\Delta = p_2-p_1$, $x$ is the average momentum fraction along the lightcone of the struck quark, the skewness $\xi = - \frac{\Delta^+}{P^+}$ encodes the momentum transfer along the light cone and $t = \Delta^2$ is the total momentum transfer squared.   

Polarised and transversity GPDs can be defined in a similar way, using the adequate operators and tensorial decomposition (see for instance \cite{Diehl:2003ny}). 
Using gluon operators, the gluon GPDs can also be defined.
However, we will stay at the level of quark unpolarised GPDs here and won't discuss these additional distributions.

We remind the reader that GPDs are probed by DVCS, whose Compton Form Factors are expressed as a convolution between a coefficient function $T$, describing the local interaction between the probe and the struck parton, and the GPDs, encoding the long distance physics:
\begin{align}
    \label{eq:Convol}
    \mathcal{H}^q(\xi,Q^2) = \int_{-1}^1 \frac{\mathrm{d}x}{2\xi} T^q\left(\frac{x}{\xi},\frac{Q^2}{\mu^2},\alpha_s(\mu^2)\right)H^{q(+)}(x,\xi,\mu^2) \, ,
\end{align}
where 
\begin{equation}
    H^{q(+)}(x,\xi,\mu^2) = H^q(x,\xi,\mu^2) - H^q(-x,\xi,\mu^2)
\end{equation}
is the charge-even GPD. Here we omit the $t$ dependence, as it plays no role in this entire paper. $Q^2$ is the photon virtuality and $\alpha_s(\mu^2)$ is the strong running coupling evaluated at $\mu^2 \simeq Q^2$. $T^q$ is nowadays known at NNLO \cite{Braun:2020yib} and NNLP \cite{Braun:2025xlp}.
However in the present study we will stick to NLO and LP, using the expressions provided in \cite{Moutarde:2013qs}:
\begin{align}
    \label{eq:CoefficientFunction}
    T^q = C_0^q + \alpha_s(\mu^2)C_1^q + \alpha_s(\mu^2)\ln \left(\frac{Q^2}{\mu^2}\right) C_{\mathrm{Coll}}^q .
\end{align}
$C_0^q$, $C_1^q$ and $C_{\mathrm{Coll}}^q $ are known functions of $x$ and $\xi$.

In the following, we will study only the imaginary part of the amplitude $\mathcal{H}$ which is generated by the imaginary part of the coefficient function $T$.
The reason comes from the fact that the imaginary part probes the DGLAP region, which is by itself enough to fully characterise the GPDs up to the $D$-term \cite{Chouika:2017dhe,Chouika:2017rzs,Chavez:2021llq,DallOlio:2024vjv}.
This $D$-term can then be extracted from the real part of the amplitude through the dispersion relations \cite{Teryaev:2005uj,Anikin:2007yh,Diehl:2007jb,Dutrieux:2024bgc,Martinez-Fernandez:2025rcg,  Kumericki:2007sa}, but faces its own deconvolution problem \cite{Dutrieux:2024bgc}.

In order to extract GPDs from the experiment, it is critical to implement as many theoretical constraints as possible, in order to mitigate the sparsity problem and tame the deconvolution one.
In this paper, we will mostly use three of them.
First, in the forward limit, the GPDs are reduced to the standard PDF following:
\begin{align}
    \label{eq:ForwardLimit}
    H(x,0,0) = q(x) \theta(x) -\bar{q}(-x)\theta(-x) .
\end{align}
The second one is that their scale dependence is known perturbatively following (see for instance \cite{Bertone:2022frx}):
\begin{align}
    \label{eq:evolution}
    H^q(x,\xi,\mu^2) = \int_{x}^\infty \frac{\mathrm{d}y}{y} \Gamma^{qi}\left(y,\frac{\xi}{x},\alpha_s;\mu^2,\mu_0^2\right) H^i\left(\frac{x}{y},\xi,\mu_0^2\right)
\end{align}
where $\Gamma$ is the GPD evolution operator.
Finally, we use the polynomiality property \cite{Ji:1998pc,Radyushkin:1998bz} that yields:
\begin{align}
    \label{eq:Polynomiality}
    \int \mathrm{d}x x^n H(x,\xi) = \sum_{j}^{\left[\frac{n}{2}\right]}A_{2j;n}(2\xi)^{2j} + \mathrm{mod}(2,n)(2\xi)^{n+1}C_{n+1},
\end{align}
where the $A$ are the generalised form factors, and the $C$ the moments of the so-called Polyakov-Weiss $D$-term \cite{Polyakov:1999gs}.

Note that in the following we will not exploit the positivity properties \cite{Pire:1998nw,Radyushkin:1998es,Diehl:2000xz,Pobylitsa:2001nt}.
An assessment of their impact has been for instance performed in \cite{Dutrieux:2021wll}.

\subsection{Double Distribution representation}

In order to consistently implement the polynomiality property, we turn to the Double Distribution (DD) representation, which is exactly equivalent \cite{Chouika:2017dhe}. 
In the DD representation \cite{Mueller:1998fv,Radyushkin:1997ki}, the GPD $H$ is given as:
\begin{align}
    \label{eq:DD}
    H(x,\xi) = & \int_{\Omega} \textrm{d}\beta \textrm{d}\alpha \delta(x-\beta -\alpha \xi) \left[F(\beta,\alpha) +   \xi \delta(\beta) D(\alpha)\right] ,
\end{align}
where $\Omega$ is the L1 ball, \emph{i.e.} $\Omega = \{(\beta,\alpha) | |\beta| + |\alpha| \le 1 \} $ .
A similar expression can be obtained for the other GPDs.
We will focus on the DD $F$ in the following, and leave the Polyakov-Weiss $D$-term for future studies.
We note that, in the DD formalism, the PDF is obtained by integrating the DD:
\begin{align}
    \label{eq:PDFfromDD}
    q(x) = \int_{-1+x}^{1-x} \mathrm{d}\alpha F(x,\alpha), \quad \mathrm{for~} x\ge 0.
\end{align}
In the following, we will thus determine how well one can reconstruct the DD using experimental data.


\section{Discretising Double Distributions}
\label{sec:Discretise}
\subsection{Mesh generation and interpolation}

Our strategy to extract the DD from experimental data is to rely on Finite Element Methods, which have been extensively studied in the past in order to reconstruct GPDs from the knowledge of the DGLAP region only (see \cite{Chouika:2017dhe,Chouika:2017rzs,Chavez:2021koz,Chavez:2021llq,DallOlio:2024vjv,Castro:2025rpx}), but also more recently in the extraction of DD from lattice QCD data \cite{Dutrieux:2026grg}.

\subsubsection{Symmetries and problem reduction}

\begin{figure}
    \centering
    \includegraphics[width=0.8\linewidth]{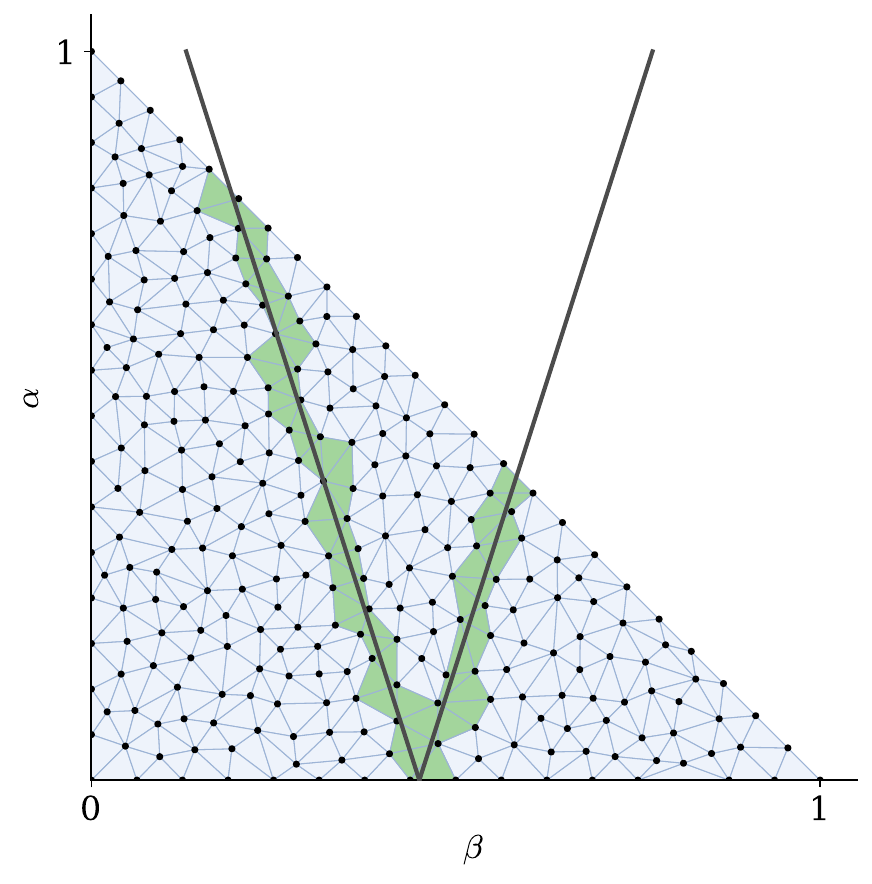}
    \caption{A Delaunay mesh with P1 interpolation nodes probed by a Radon line.}
    \label{fig:UpperrightSupport}
\end{figure}

The first step consists in discretising the DD support domain. In that regard, two things can be mentioned. 
First, (and for a reason that will become clear in Sec. \ref{sec:nonlocal}), we want to avoid sliver triangles, so that, even if the mesh itself is irregular, the triangulation does not present pathologies.
This pushes us toward a specific type of triangulation called the Delaunay triangulation, which is specifically designed in order to avoid sliver triangles.
In what follows, we will thus rely on this algorithm to build the mesh, and discuss only one parameter, $a$, which is the maximal area authorised for a triangle.

That said, we do not want to triangulate the entire domain $\Omega$.
Indeed, we know that time-reversal invariance, parity and hermiticity together ensure that the GPDs $H$ and $E$ are even in $\xi$, which can directly be translated in terms of the DD $F$ to be even in $\alpha$.
We therefore restrict ourselves to the upper plane $\alpha >0$.

Furthermore, the combination of $s$ and $u$ channels in the DVCS description selects the $x$-odd component of the GPDs (as shown in Eq. \eqref{eq:Convol}, depending only on $H^{q(+)}$). 
Since $F$ is even in $\alpha$, the odd-parity in $x$ of the GPDs in $(x,\xi)$ space is recovered through the odd-$\beta$ part of the DD distribution $F$, that we note $F^{(+)}$. And since $F^{(+)}$ is odd, we can reduce the current study simply to the upper right quadrant $\{(\beta,\alpha) | \beta >0 \mathrm{~and~} \alpha>0\}$.
In this quadrant, the discretised support of the DD is presented in Fig. \ref{fig:UpperrightSupport}.

\subsubsection{Interpolation degrees and shadow modes}
\label{sec:Interpolation}

Once the mesh has been generated, one needs to define a low-degree polynomial interpolation, independent on each element.
The coefficients of the polynomials are fixed by the values of the DD at nodes, whose position on each element depends on the polynomial degree.
We will respectively denote by P0, P1 and P2 the local polynomial interpolation of degree 0, 1 and 2 on each element.
For the P0 case, a single node is necessary, located at the barycentre of each triangle.
This interpolation is constant by pieces, and discontinuous.
For P1, three nodes are required, located at the vertices of the mesh, \emph{i.e.} at the corner of each triangle.
The nodes are thus shared between multiple elements, ensuring that the interpolation is continuous, and planar by pieces.
For P2, six nodes are necessary, located at the corners of each triangle and at the middle of the edges. 
The interpolation is still continuous, but the number of degrees of freedom is not enough to ensure the continuity of the first derivative. 
We expect of course that most of the time\footnote{We will see in Sec. \ref{sec:LO_convergence} that it is not always the case.} increasing the interpolation order will increase the convergence toward the targeted DD.

However, one cannot target a polynomial interpolation of arbitrarily high degree on every element for the problem we are interested in, reconstructing the DD from experimental data.
As it was shown in Ref. \cite{Bertone:2021yyz}, one can find polynomial DDs that belong to the null space of DVCS, called shadow DDs.
More precisely, at leading order, Ref. \cite{Bertone:2021yyz} highlighted a family of degree 9 and above polynomial DDs, such that their forward limits vanish and the associated GPDs $H(\xi,\xi)$ also vanish.
However, the maximal polynomial degree that one can use in an LO treatment of the DVCS shadow DD determination is in fact much smaller.
Indeed, Ref. \cite{Dutrieux:2022snd} shows that a DD written as an order 5 polynomial, even in $\alpha$ and odd in $\beta$, is already enough to cancel the LO CFF.
But even that case is overestimating the maximal polynomial order one can pick at LO in our proposed work. 
The polynomial shadow DDs computed in \cite{Dutrieux:2022snd} are smooth functions, defined on the entire DD domain.
Here, we define the DD on the upper-right quadrant of the $\mathbb{R}^2$ plane, and reconstruct the full DD by symmetrisation in $\alpha$ and antisymmetrisation in $\beta$.
The direct consequence is that we allow for DDs which are discontinuous on the $\beta =0$ line, and whose first derivative is discontinuous on the $\alpha = 0$ line.
We therefore probe a significantly larger functional space than the one spanned in \cite{Bertone:2021yyz,Dutrieux:2022snd}.

Within the functional space probed, the first LO partial shadow mode\footnote{By partial shadow mode, we mean that only the condition of belonging to the null space of DVCS is realised, but not the one of vanishing forward limit.} is found for a degree as low as 2 : 
\begin{equation}
    \label{eq:PartialShadowDDLO}
    F_{0}^q(\beta,\alpha) =  \lambda \left( \beta^2 - \alpha^2 -\frac{2}{3}\beta +\frac{1}{3}\right), \quad \mathrm{for~} \beta>0,~ \alpha >0, 
\end{equation}
where $\lambda$ is an arbitrary normalisation constant.
Interestingly, Eq. \eqref{eq:PartialShadowDDLO} is the Nakanishi-based model introduced in Ref. \cite{Chouika:2017rzs}.
It is thus impossible to use meaningfully a P2 approximation for extraction of the DD at LO, or at least without using other constraints such as the forward limit to regulate the partial shadow mode highlighted above.
In order to avoid this complication, when dealing with the DVCS LO approximation, we will simply restrict ourselves to the P0 and P1 interpolation.
Note that at NLO, this constraint disappears and one can safely use a P2 approximation. 

\subsubsection{PDF bias and non-local tessellations}
\label{sec:nonlocal}

Before moving on to testing the reconstruction of the DD itself, we need to address an important question: how strong are the information coming from the PDF in a DD extraction?
Assuming that the PDF is known with high precision, one could imagine sampling it at will in order to probe, through integration over $\alpha$ as in Eq. \eqref{eq:PDFfromDD}, all cells of our discretisation, multiple times.
The question thus boils down to assessing whether the forward limit can strongly bias a genuine off-forward extraction, and how.

\begin{figure}
    \centering
    \includegraphics[width=0.95\linewidth]{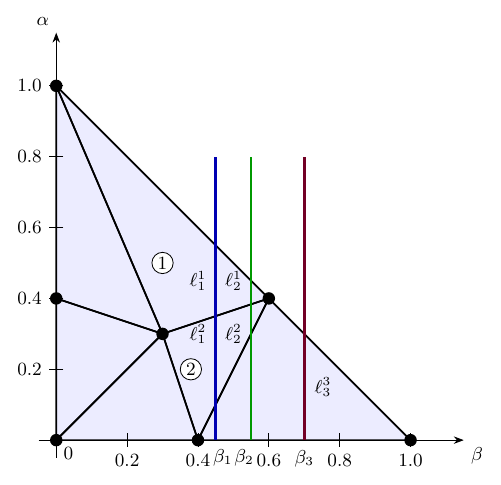}
    \caption{PDF-type lines along which the DD is integrated. Even for the most simple case of P0 interpolation, the geometry of the system forbids an independent determination of the DD on cells 1 and 2 from PDF lines only.}
    \label{fig:PDFBias}
\end{figure}

A simple, geometric answer to this question is provided in Fig. \ref{fig:PDFBias}.
On a standard, local, Delaunay mesh, the PDF-type lines cross multiple cells along their path.
However, they are all parallel one with the other.
For a P0 interpolation on each cell, the result of the integration is simply given by the length of the path $\ell$ of a given line $i$ through the cell $e$, multiplied by the constant $c_e$ approximating the DD on the cell $e$.
For Fig. \ref{fig:PDFBias}, it yields the following system:
\begin{equation}
  \label{eq:PDFSystem}
  \left\{
    \begin{array}[h]{rcl}
      c_1 \ell_1^1 + c_2 \ell_2^1 + c_3 \ell_3^1 & = & q(\beta_1) \\
      c_1 \ell_1^2 + c_2 \ell_2^2 + c_3 \ell_3^2 & = & q(\beta_2)\\
      c_3 \ell_3^3 & = & q(\beta_3)
    \end{array}
    \right. ,
\end{equation}
which is not invertible since
\begin{equation}
  \label{eq:ThalesTheorem}
  \frac{\ell_2^1}{\ell_1^1} = \frac{\ell_2^2}{\ell_1^2}\, ,
\end{equation}
by Thales's intercept theorem.
The geometry of the problem forbids us to extract independently $c_1$ and $c_2$, and therefore, the PDF cannot determine the off-forward physics. 

\begin{figure}
    \centering
    \includegraphics[width=0.95\linewidth]{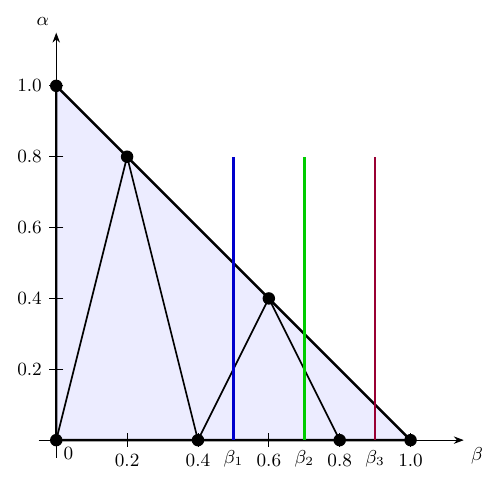}
    \caption{Non-local geometry leading to an invertible FEM system by the knowledge of the PDF only.}
    \label{fig:PathologicalPDF}
\end{figure}

One can, however, figure out a geometry in which the system remains invertible.
An example is provided in Fig. \ref{fig:PathologicalPDF}.
This apparent paradox is solved by noting that the triangulation of Fig. \ref{fig:PathologicalPDF} is non-local, \emph{i.e.} when the area of the triangles goes to zero, the size of the edges does not, and the system yields more and more sliver triangles. 
In the limit $a\to 0$, they span an infinitesimal $\mathrm{d}\beta$ range, while averaging the DD on the entire range of $\alpha$.
Consequently, off-forward information is also lost, and the apparent paradox is solved.

\subsection{Benchmarking on GK}
\label{sec:GKBenchmark}

For the purposes of this study, a Python code was developed and benchmarked on the simplified Goloskokov-Kroll (GK) DD model \cite{Goloskokov:2007nt}
\begin{align}
    F_i(\beta, \alpha, t)&=2N |\beta|^{-\alpha_i(t)}(1-|\beta|)^{b_i} \times\nonumber\\
    &\times \frac{\Gamma\left(2 n_i+2\right)}{2^{2 n_i+1} \Gamma^2\left(n_i+1\right)} \frac{\left[(1-|\beta|)^2-\alpha^2\right]^{n_i}}{(1-|\beta|)^{2 n_i+1}} \Theta(\beta),\label{GK}
\end{align}
with
\begin{equation}
\alpha_i=\alpha_i(0)+\alpha_i^{\prime} t, \quad i=g, \text { sea, val }.
\end{equation}
Having in mind to first study the non-singlet sector of GPDs through DVCS (such as $H^{u-d}$), we pick the ``valence" parameters of the GK model, namely 
$n_{\text{val}}=1$, $b_{\text{val}}=3$, $\alpha_{\text{val}}(0)=0.48$ and $\alpha'_{\text{val}}=0.9$, with $N$ fixed by 
\begin{equation}
    \int_0^1\mathrm{d} x\, xH(x,0)=\frac{1}{3}.
\end{equation}
Since DVCS probes the charge-even GPD combination $H^{q(+)}$, we antisymmetrise the reconstructed GPD and use this for all analysis in this work. We use $t=0$ throughout, unless otherwise specified.

\subsubsection{Reconstruction of GK model}

\begin{figure}
    \centering
    \includegraphics[width=0.9\linewidth]{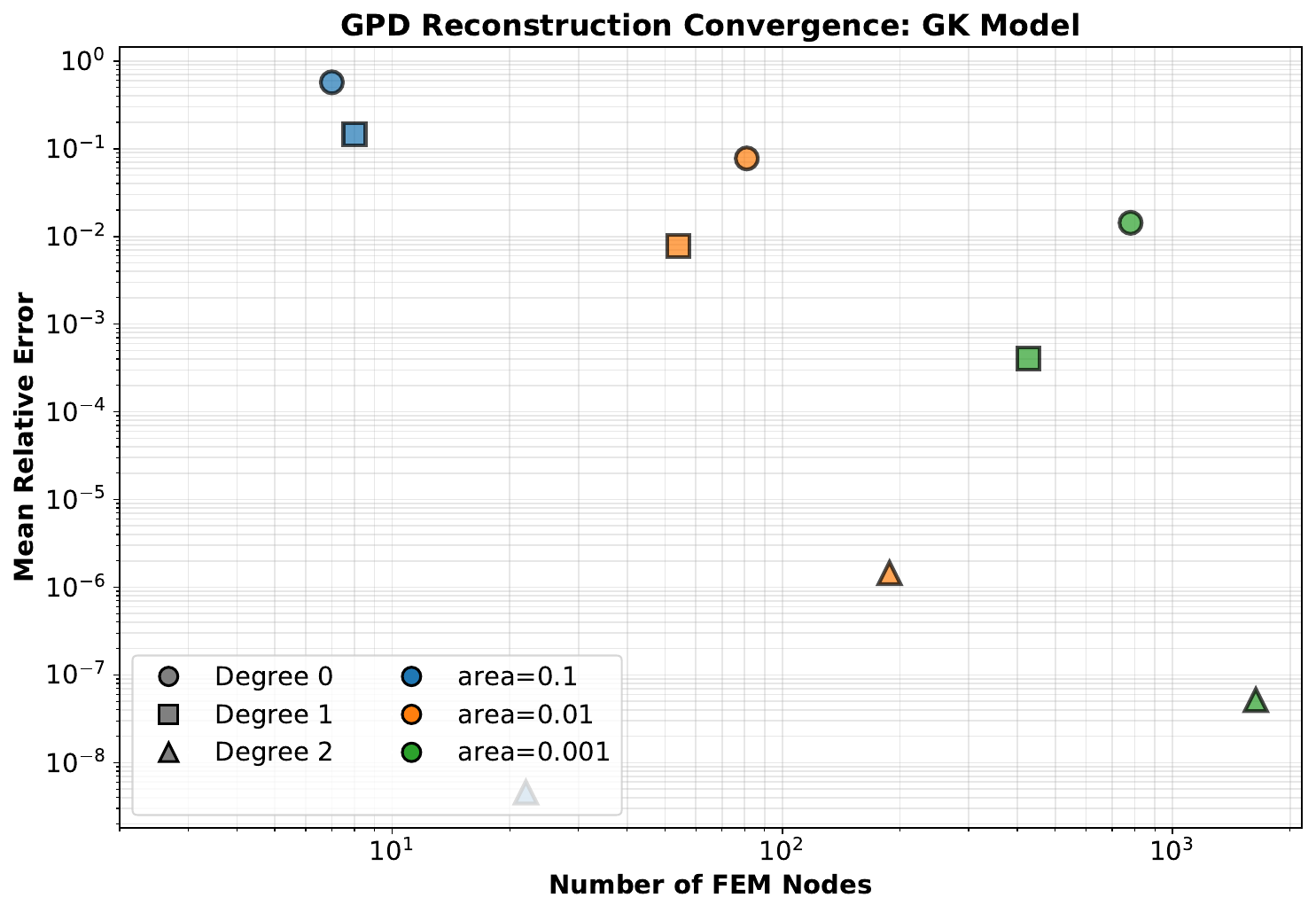}
    \caption{Reconstruction error of the charge-even GK GPD with respect to mesh size and interpolation order.}
    \label{fig:gk_rec}
\end{figure}

Due to the divergent behaviour of Eq. (\ref{GK}) at $\beta=0$, we implement a hybrid integration routine. Namely, for all matrix elements that sample mesh elements touching $\beta=0$, there is an end-point singularity in the integral, symbolically written as
\begin{equation}
   \int_0^{\beta^{out}}\mathrm{d}\beta \,\beta^{-\alpha_{\mathrm{val}}(0)} f_{\mathrm{finite}}(\beta),
\end{equation}
where $\beta^{out}$ is the value of $\beta$ where the sampling line exits the sampled element, which is resolved with the Gauss-Jacobi quadrature. Away from $\beta=0$, the weight $\beta^{-\alpha_{\mathrm{val}}(0)}$ is included in the integrand of a standard Gauss-Legendre quadrature, meaning that the FEM only interpolates the polynomial part of the GK DD. In Fig. \ref{fig:gk_rec}, we plot the mean relative error of reconstructing the GK GPD with varying polynomial interpolation, where we observe the expected jump in precision once we use P2 interpolation, since the Regge-stripped GK DD chosen here is a quadratic polynomial in both $\beta$ and $\alpha$.

\subsubsection{Impact of the Regge uncertainties}

\begin{figure}
    \centering
    \includegraphics[width=0.95\linewidth]{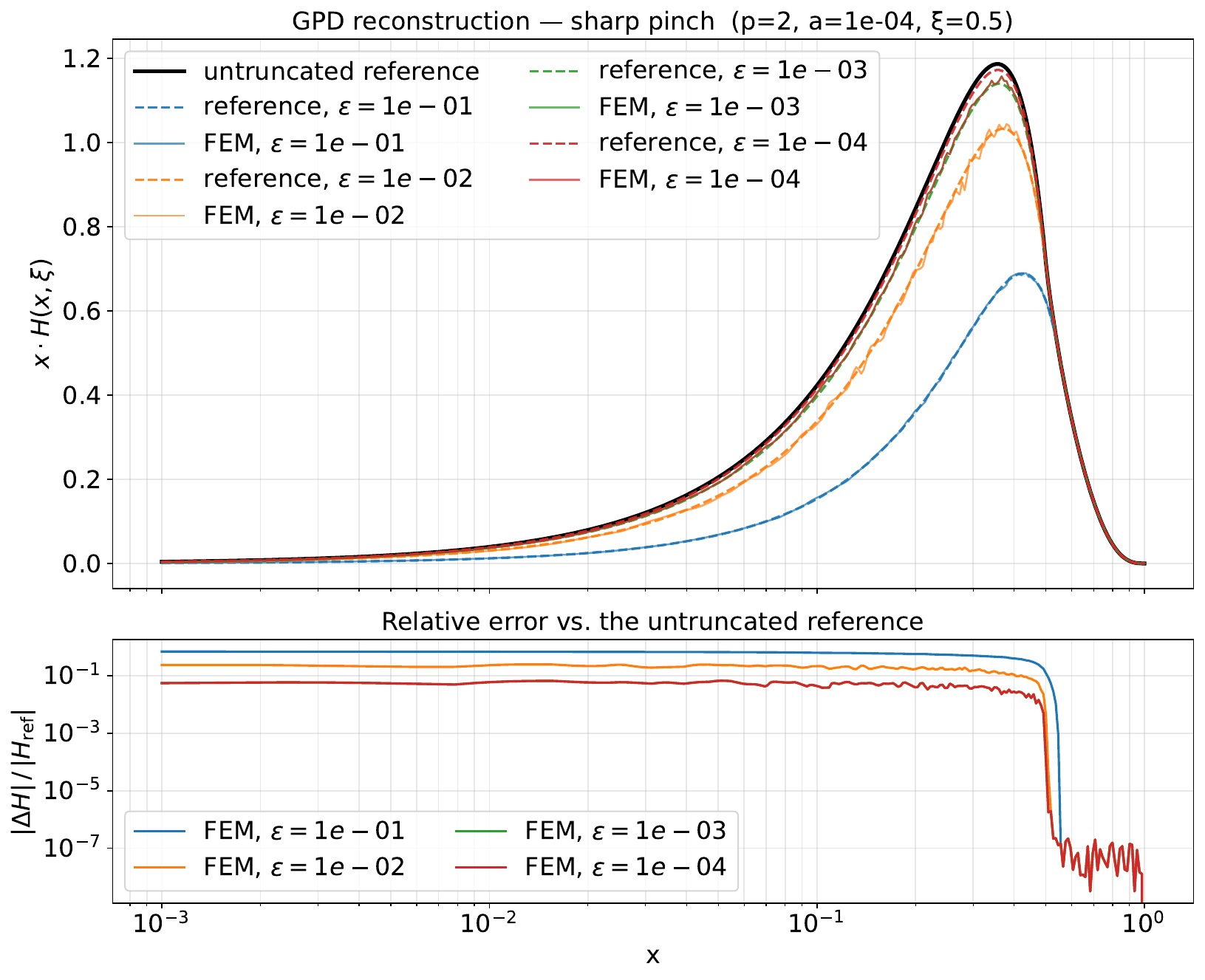}
    \caption{Reconstruction of the GK-like GPD with mismatched Regge weight, truncated at $\beta<\varepsilon$ for $\varepsilon=10^{-1}$ to $10^{-4}$, with P2 interpolation and maximal element area $10^{-4}$. Top: $xH(x,\xi=0.5)$ from the FEM (full lines) and from direct numerical integration of the truncated DD (dashed); the black line is the untruncated GPD. Bottom: relative error of the FEM with respect to the untruncated GPD.}
    \label{fig:Pinch_GK}
\end{figure}

\begin{figure}
    \centering
    \includegraphics[width=0.95\linewidth]{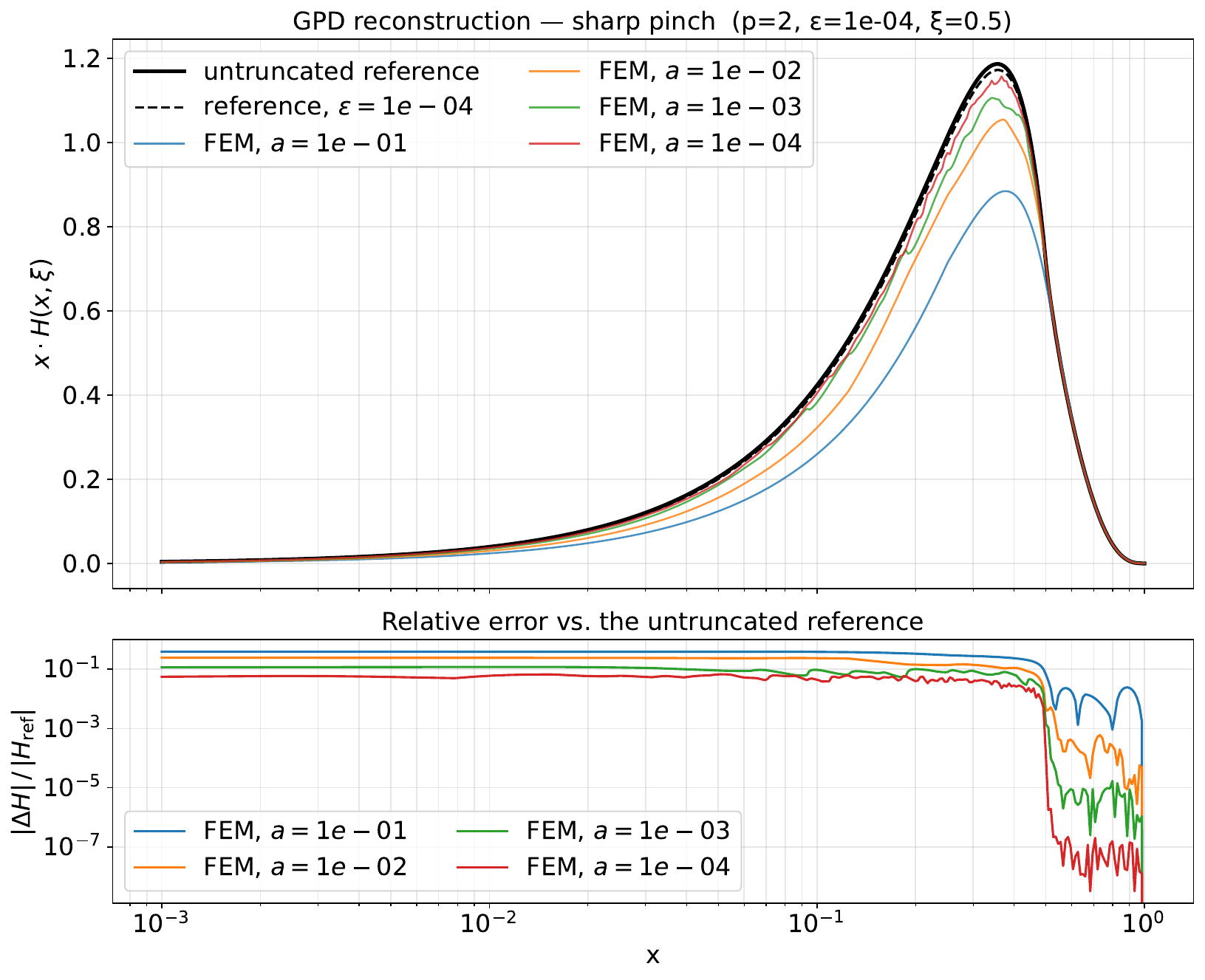}
    \caption{Same as Fig.~\ref{fig:Pinch_GK} at fixed $\varepsilon=10^{-4}$, for maximal element areas $a=10^{-1}$ to $10^{-4}$. In the top plot, we show how the FEM reconstruction of $xH(x,\xi=0.5)$ slowly converges to the expected results when $a\to 0$, highlighting the important impact of uncertainties on the Regge behaviour on the ERBL reconstruction.}
    \label{fig:Pinch_GK_FEM}
\end{figure}

While it is safe to assume the presence of a Regge pole in any experimentally viable extraction of DDs, based on some of the most recent extractions of PDFs \cite{Ball2017asymptotic, NNPDF2022, CT18, MSHT20, PDF4LHC21}, hard-coding a weight could pose problems in future data analyses, where the goal would be to extract a DD from an experimentally constrained GPD, in which case we do not know the exact power of the Regge pole. In order to estimate the potential error of not knowing the exact power, we take a conservative approach by setting $\alpha_{w}=0.99$ instead of $\alpha_{\mathrm{val}}(0)=0.48$ and reconstructing
\begin{equation}
    G(\beta,\alpha)=\beta^{0.51}F(\beta,\alpha).
\end{equation}
Since this is not an integer-order polynomial, the precision of the reconstruction only mildly\footnote{By mildly, we mean here that the power law of the precision reached is not impacted by the polynomial order, only the prefactors are.} improves with higher-order interpolation.
This problem is formalised with the Babuška-Suri error estimate \cite{Babuska}, which shows that the convergence rate of an approximation of a function with a corner singularity of the type $\beta^s$, $s\notin \mathbb{Z}$, by degree-$p$ finite polynomials on a quasi-uniform mesh saturates as $h^{s+1/2}$, where $h$ is the dimension of the mesh element, and is independent of the polynomial degree $p$ when $s+1/2<p+1$. 
In this case, with $s=0.51$, the convergence rate saturates at $p\geq1$ and is only approximately linearly improved by reducing the size of the mesh elements.
In an effort to characterise this problem, we regularise the singularity $\beta=0$ by placing a cut-off $\varepsilon$ on the bottom boundary of the Radon integral. 
This cutoff has two aims.
First, it allows us to estimate how quickly the Radon integrals converge when taking $\varepsilon \to 0$ (see Fig. \ref{fig:Pinch_GK}). Then, it allows us to assess how fine the mesh should be in order to properly reconstruct the impact of the Regge behaviour on the ERBL region (see Fig. \ref{fig:Pinch_GK_FEM}).
The error incurred by this method is estimated as
\begin{equation}
   \int_0^{\varepsilon}\mathrm{d}\beta\, \beta^{-\alpha_{\mathrm{val}}(0)} f_{\mathrm{finite}}(\beta)\approx \frac{f_{\mathrm{finite}}(0)}{1-\alpha_{\mathrm{val}}(0)}\varepsilon^{1-\alpha_{\mathrm{val}}(0)},
\end{equation}
and is of order 1\% for $\varepsilon=5\times10^{-5}$. In Fig. \ref{fig:Pinch_GK}, we show the comparison between a pinched and unpinched Radon transform and the error of the FEM reconstruction with such a pinch relative to the unpinched Radon transform calculated by direct numerical integration. We see that the red and green FEM lines coincide fully, as the reconstruction saturates at $\varepsilon=10^{-3}$ at maximum element area $10^{-4}$. Due to the GPD having small values as $x\to1$, we floor the denominator in the calculation of the relative error to 1\% of the maximum value of the GPD.

Figs. \ref{fig:Pinch_GK} and \ref{fig:Pinch_GK_FEM} highlight the importance of getting the Regge behaviour of the PDF right in order to properly describe the ERBL region.
This result may seem counter-intuitive.
Indeed, the ERBL region is usually presented as the ``meson-like" kinematic region of the GPDs, and is nevertheless strongly impacted by the Regge behaviour of the PDF.
As a consequence, it is probably better to build a DD model ``à la Radyushkin" \cite{Musatov:1999xp}, in which the PDF is taken as an input of the DD modelling rather than trying to extract the whole DD.
In the latter case, we can foresee an inflated uncertainty in the ERBL region coming from the additional uncertainty associated with the extraction of the Regge behaviour.
The determination of the Radyushkin profile function is at first sight the safest choice.


\section{The deconvolution problem}
\label{sec:Deconvolution}

\subsection{FEM reconstruction of shadow GPDs}

As already highlighted in Eq. \eqref{eq:Convol}, the CFFs parameterising the Compton amplitudes are given in terms of a convolution between our targeted GPDs and a coefficient function $T$.
Unfortunately, this convolution is not invertible at fixed scale: the associated null space is non-trivial and elements of the null space, called shadow GPDs, satisfying the known physical constraints can nevertheless significantly contaminate the extraction of GPDs \cite{Bertone:2021yyz}.

Shadow GPDs by construction must also satisfy polynomiality, which means they can be connected to shadow DDs. In \cite{Bertone:2021yyz} it was shown that one can recover a family of polynomial shadow DDs at LO and NLO, which can approximate any continuous DD to arbitrary precision. In this work, we take the LO and NLO shadow DDs from this paper and study their contamination of the GPD in the FEM. Since FEM has finite precision, we expect that the shadow contributions will not vanish identically in the CFF and PDF, and we also expect that these contributions will tend towards zero as we refine the FEM discretisation. 

\subsubsection{LO with P0 and P1 approximation}
\label{sec:LO_convergence}

At LO, we take a $9^{\text{th}}$-order polynomial, iso-vector shadow DD, construct the corresponding GPD using the FEM at different levels of precision and with different interpolation, and use this GPD to calculate the PDF and the LO CFF. Since shadow DDs belong to the DVCS and DIS null space, their normalisation is arbitrary. In this subsection, we conveniently fix the normalisation of all shadow GPDs so that the maximum value of $|H^{\text{+,shadow}}(x,\xi=0.5)|$ in the DGLAP region is 1. 

\begin{figure}
    \centering
    \includegraphics[width=0.9\linewidth]{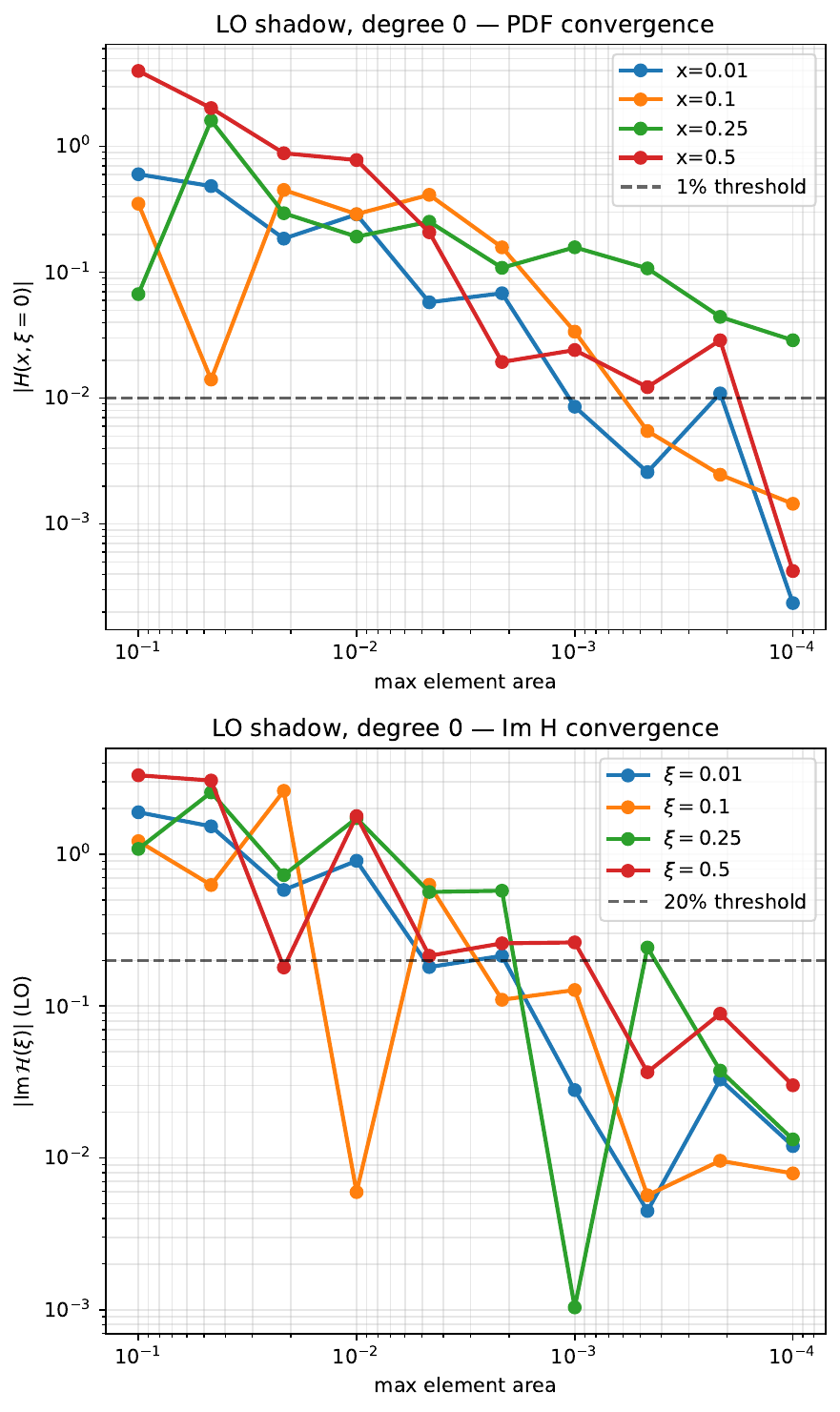}
    \caption{Reconstruction of the LO shadow PDF and $\mathfrak{Im}\mathcal{H}$ with P0 interpolation with increasing mesh refinement.}
    \label{fig:LO0}
\end{figure}

\begin{figure}
    \centering
    \includegraphics[width=0.9\linewidth]{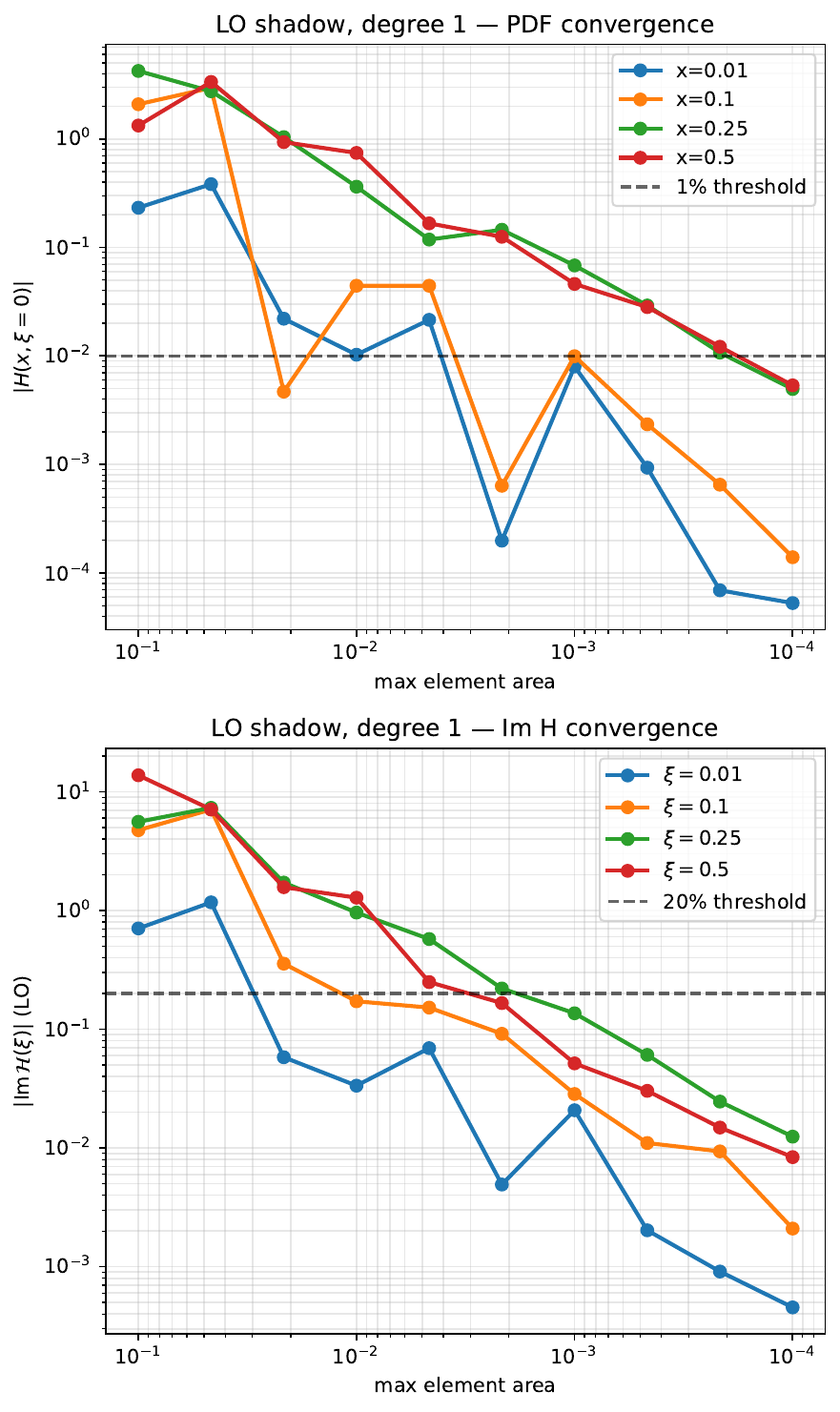}
    \caption{Reconstruction of the LO shadow PDF and $\mathfrak{Im}\mathcal{H}$ with P1 interpolation with increasing mesh refinement.}
    \label{fig:LO1}
\end{figure}

In Figs. \ref{fig:LO0} and \ref{fig:LO1}, we observe how the shadow PDF $|H(x,\xi=0)|$ and CFF $|\mathfrak{Im}\mathcal{H}(\xi)|$ decrease in value with mesh refinement, represented by the maximum element area $a$, at $x$ and $\xi=0.01,0.1,0.25,0.5$, which is further improved by going from P0 to P1 interpolation. The aim is to find $a$ such that the shadow DD contaminates the PDF by less than 1\% and the CFF by less than 20\%, which is the case with P1 interpolation.

The result shows that surprisingly, the considered shadow GPD is too roughly approximated by a mesh as fine as $a=10^{-3}$ to be invisible in the kinematic valence region $\xi \in [0.1,0.5]$.
Another unexpected outcome is the relatively better assessment of the shadow DD for a P0 interpolation compared to a P1 one at large $a$.
This can be understood by a stronger edge effect in the P1 case, reinforced by the fact that the nodes of the interpolation polynomials lie at the vertices of the triangle for P1, while they are at the barycentre for P0.
Therefore, the P1 interpolation is more sensitive to what happens at the edges, and, when the triangulation is coarse, edge effects are carried deeper within the support domain.
This explains why when $a=0.1$ (i.e. a number of elements strictly smaller than 10), the P0 approximation yields a better result.

One should nevertheless highlight that Figs. \ref{fig:LO0} and \ref{fig:LO1} are a single example of a function belonging to the joint LO DVCS and DIS null space.
The reader should keep in mind the caveat that other elements of the functional null space may decrease more rapidly with $a$.

\subsubsection{NLO with P1 and P2 approximation}

\begin{figure}
    \centering
    \includegraphics[width=0.9\linewidth]{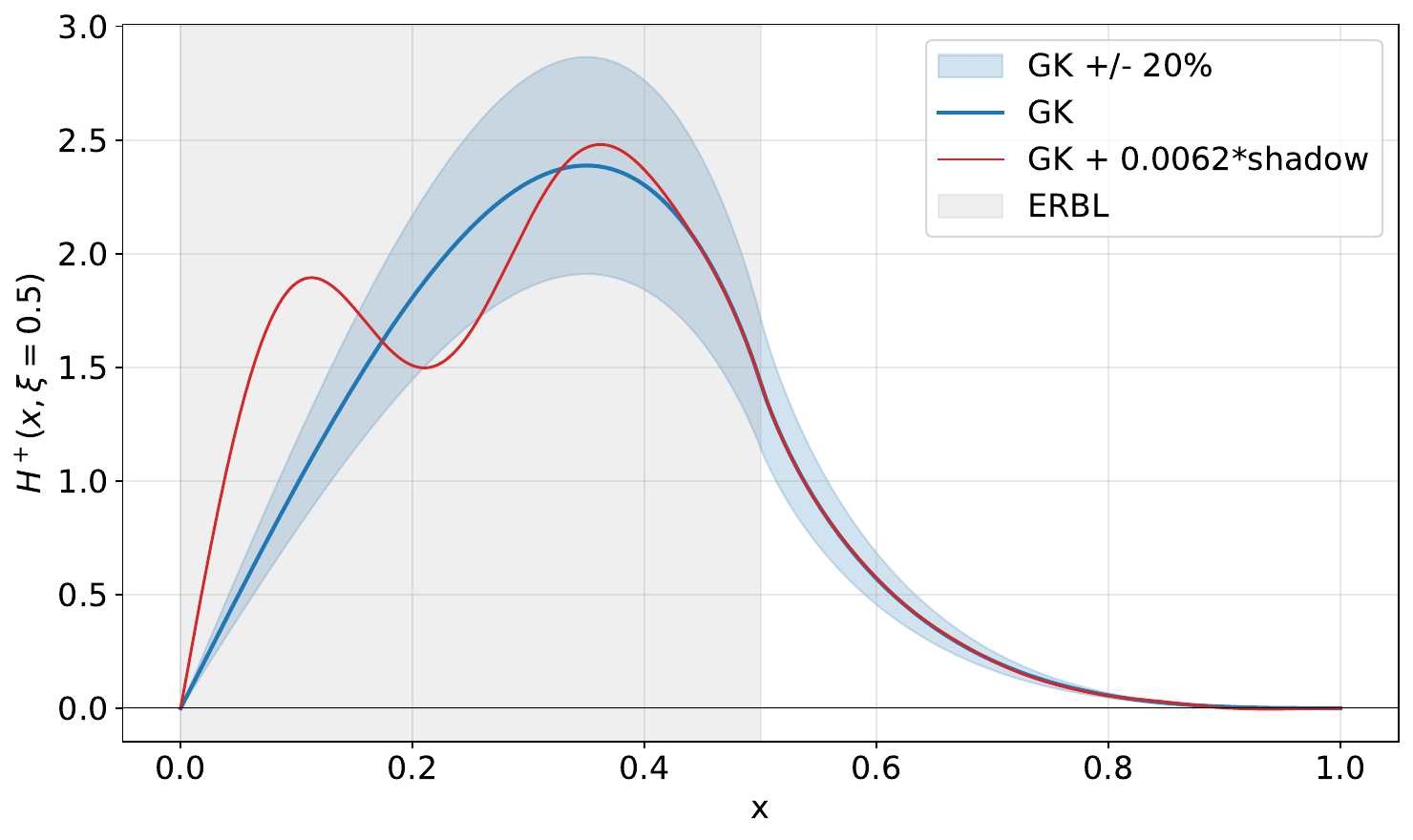}
    \caption{NLO shadow contamination of the GK GPD at $\xi=0.5$.}
    \label{fig:contamGK}
\end{figure}

At NLO, we select a combination of shadow DDs from the basis of $29^{\text{th}}$-order polynomials whose
contribution is concentrated in the ERBL region, since this region is not directly constrained by experiment
and is the most challenging one to resolve. As in the LO case, the resulting shadow GPD is
first normalised the same way as the LO case. We then rescale it by a factor $\lambda_0$, chosen such that $H^{+,\rm GK} + \lambda_0 H^{+, \rm shadow}$ stays within a 20\% band around $H^{+,\rm GK}$ for 68\% of the values of $x$ at $\xi = 0.5$. This gives $\lambda_0 = 0.0062$,
and the resulting contamination of the GK GPD is shown in Fig.~\ref{fig:contamGK}. In the following, the shadow
GPD always includes this factor, so that a rescaling $\lambda = 1$ in Sec. \ref{sec:Resolution} corresponds to a 20\% band on the GPD. We highlight that, despite being a combination of polynomials of degree 29, the shadow GPD is rather smooth, and the oscillating behaviour is strongly tamed.

The same analysis as for the LO case is repeated, using P1 and P2 polynomials to interpolate the shadow DD. Within the range of $a$ considered, P2 interpolation is required to obtain 1\% and 20\% precision on the PDF and CFF, respectively.

\begin{figure}
    \centering
    \includegraphics[width=0.9\linewidth]{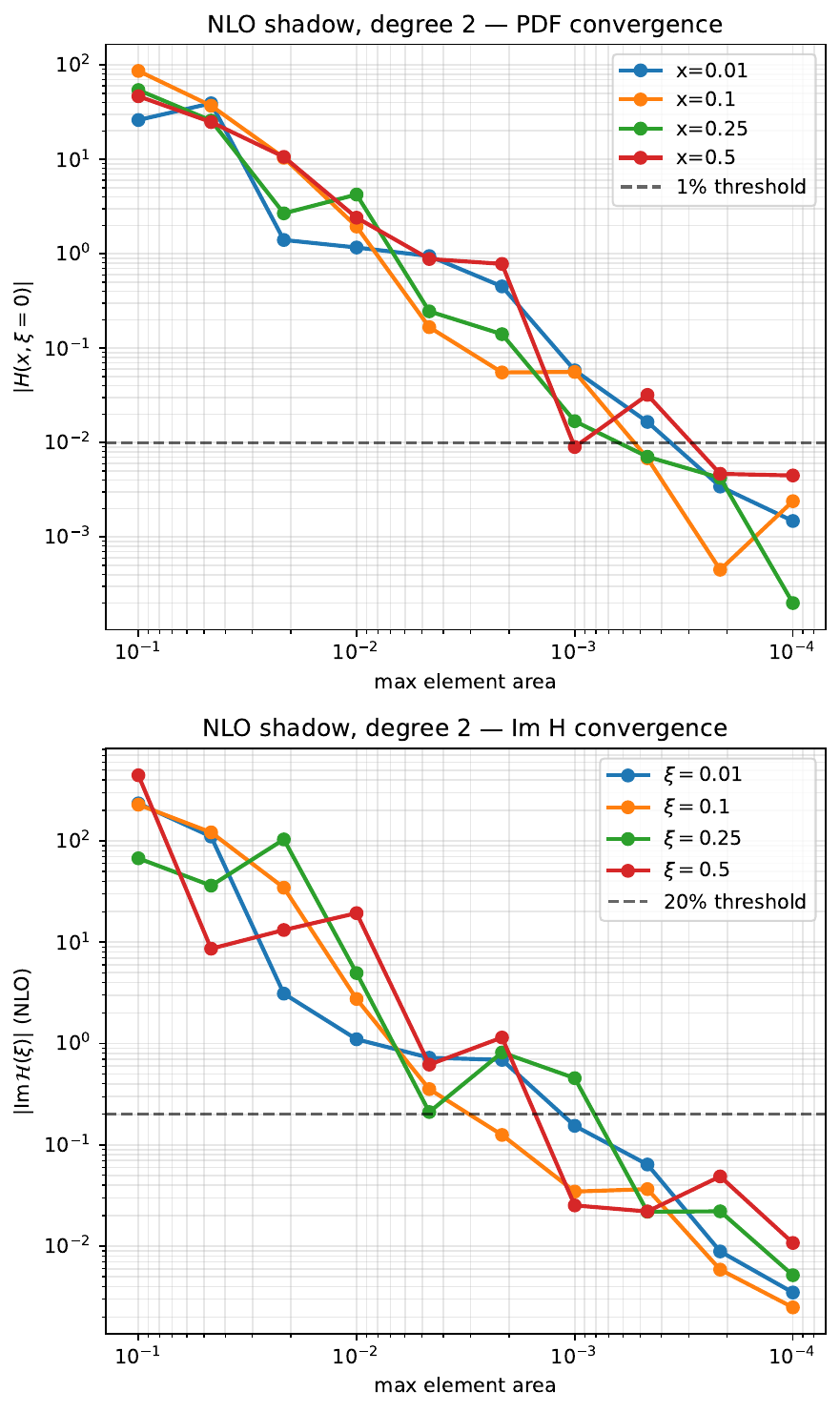}
    \caption{Reconstruction of the NLO shadow PDF and $\mathfrak{Im}\mathcal{H}$ with P2 interpolation with increasing mesh refinement.}
    \label{fig:NLO2}
\end{figure}

The results shown in Fig. \ref{fig:NLO2} need to be put in perspective with the convergence rate exemplified in Fig. \ref{fig:Pinch_GK_FEM}.
There, we have shown that even with $a=10^{-4}$, it was not possible to precisely reconstruct the behaviour in the ERBL region, and that the GPD was systematically underestimated even using P2 polynomials.
And yet, with the same polynomial accuracy and the same mesh, Fig. \ref{fig:NLO2} highlights that the shadow GPD becomes hidden in the statistical noise of the amplitude as its impact on the DVCS CFF drops below 20\%.
Stated differently, it means that the uncertainty band associated with the shadow GPD is larger than the 20\% one illustrated on Fig. \ref{fig:contamGK} and used for the normalisation of the NLO shadow GPD.
We now face a problem in which a better accuracy (i.e. our ability to reconstruct the correct answer so a smaller value of $a$) conflicts with a better precision (i.e. our ability to reduce the uncertainty band due to shadow GPDs, requiring larger value of $a$).
The optimal trade-off between precision and accuracy will depend on the quality of the experimental data, the physical constraints implemented by construction, and the pQCD precision used to describe the parton-photon scattering amplitude.

\subsection{Impact of Evolution}

\begin{figure}
    \centering
    \includegraphics[width=0.9\linewidth]{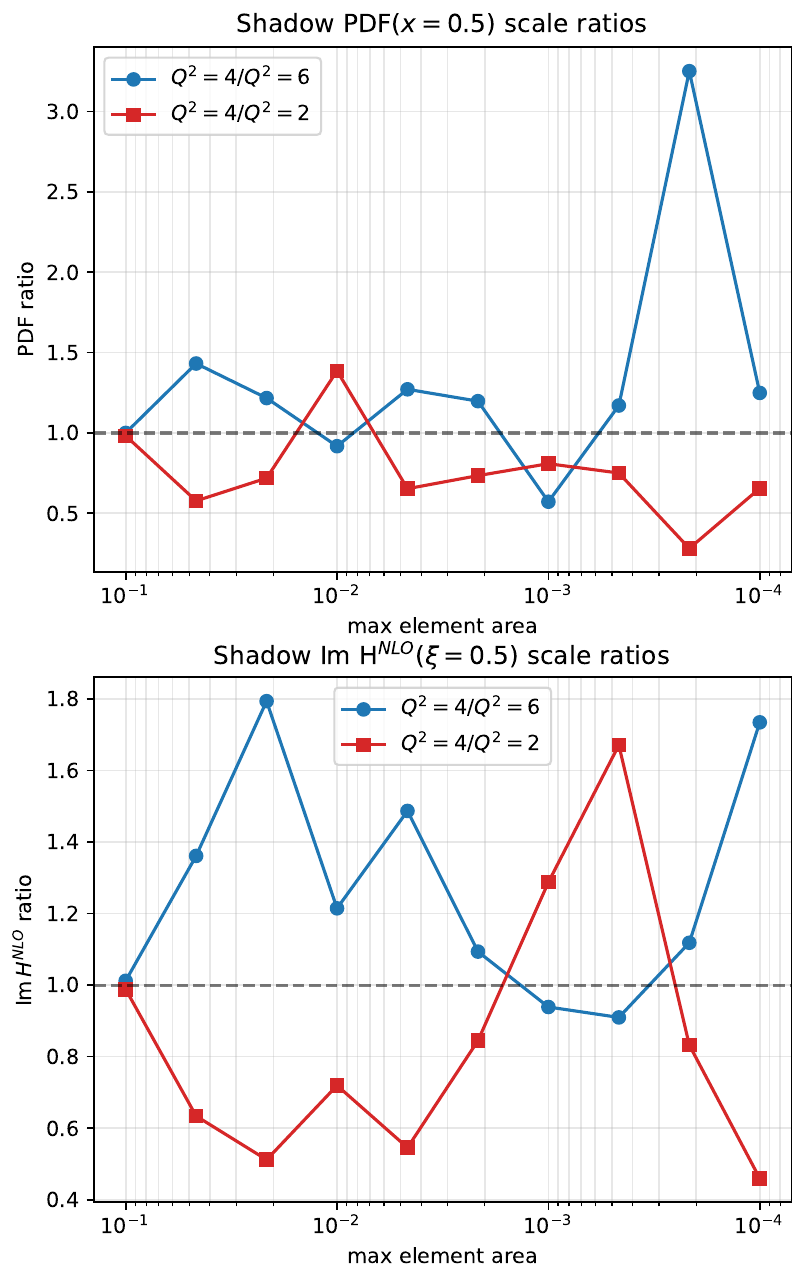}
    \caption{Convergence of the NLO shadow’s scale ratios with mesh refinement. Shown are ratios of the shadow NLO PDF at $x=0.5$ and $\mathfrak{Im}\mathcal{H}$ at $\xi = 0.5$, evaluated between the input scale $Q^2 = 4\text{ GeV}^2$ and two target scales: $Q^2 = 6\text{ GeV}^2$ (upward evolution, blue circles) and $Q^2 = 2\text{ GeV}^2$ (downward evolution, red squares), as a function of the maximum FEM element area with P2 interpolation.}
    \label{fig:Ratioevolution}
\end{figure}

The remaining question is the one of the impact of evolution on a discretised shadow GPD.
It was argued in Ref. \cite{Freund:1999xf}, that evolution allows one to recover an invertible inverse problem.
However, Ref. \cite{Bertone:2021yyz} highlighted that this improvement was in practice not meaningful, while the null-space elements do yield non-vanishing results after evolution, their impact on experimental data remains orders of magnitude below the detection threshold.
This is partly explained by the fact that the NLO shadow GPDs cancel the term behaving as $O(\alpha_s)$ in the perturbative expansion of the evolution operator (see for instance \cite{Bertone:2021yyz,Dutrieux:2022snd}).
Still, this conclusion holds only for the exact null-space elements, not their discretised approximation, which can itself yield well-detectable results at low resolution.
We thus investigate here the impact of LO evolution both forward from 4 to 6 $\mathrm{GeV}^2$ and backward from 4 to 2 $\mathrm{GeV}^2$ (see Fig. \ref{fig:Ratioevolution}).
The results do not provide a clear picture. Selecting an initial scale of $Q_0^2 = 4\, \mathrm{GeV}^2$, we compute the ratios $\mathfrak{Im} \mathcal{H}(Q_0^2) /\mathfrak{Im} \mathcal{H}(Q^2)$ for $Q^2 \in \{2\, \mathrm{GeV}^2,6\, \mathrm{GeV}^2 \}$ allowing us to probe the effect of forward and backward evolution.
An improvement of the deconvolution problem at a fixed value of $a$ would translate into a ratio smaller than one, which is not unambiguously observed.
We can probably only conclude that the discretisation effects are much stronger than the evolution effects, consistently with the discussion in Ref. \cite{Bertone:2021yyz}. 

\subsection{Resolution power of DVCS}
\label{sec:Resolution}
The FEM description of DDs allows us to bypass the deconvolution problem of DVCS by introducing $a$, the maximal area of a given element, as a regulator parameter, turning a non-invertible problem into an ill-conditioned one.
In order to characterise the impact of the null space on the extraction of the GPD, we now introduce a new procedure in which the experimental uncertainty is fixed to 20\%. 
We then introduce a rescaling factor $\lambda$ to the shadow GPD previously considered, such that when discretised with resolution $a$, the impact of the shadow GPDs remains in the mock experimental uncertainty band, \emph{i.e.}:
\begin{equation}
    \label{eq:LambdaCondition}
    |\lambda(a)\mathfrak{Im} \mathcal{H}_{\mathrm{shadow}}(a)| \le  0.2 |\mathfrak{Im} \mathcal{H}_{\mathrm{GK}}| ,
\end{equation}
where $\mathfrak{Im} \mathcal{H}_{\mathrm{GK}}$ is the imaginary part of the CFF generated by the GK model, while $\mathfrak{Im} \mathcal{H}_{\mathrm{shadow}}(a)$ is the imaginary part of the discretised shadow CFF. 
Eq. \eqref{eq:LambdaCondition} can be strictly enforced, yielding $\lambda_\mathrm{strict}(a)$ or statistically enforced (\emph{i.e.} for 68\% of the definition domain in $\xi$) yielding $\lambda_{68}(a)$.
In both cases, $\lambda(a) = 1$ corresponds to our original normalisation presented in Fig. \ref{fig:contamGK}.\\
We then compute $\lambda_{68}$ and $\lambda_{\mathrm{strict}}$ and the 68\% uncertainty band associated with the fluctuations the shadow GPD generates around the GK model.

\begin{figure}
    \centering
    \includegraphics[width=0.9\linewidth]{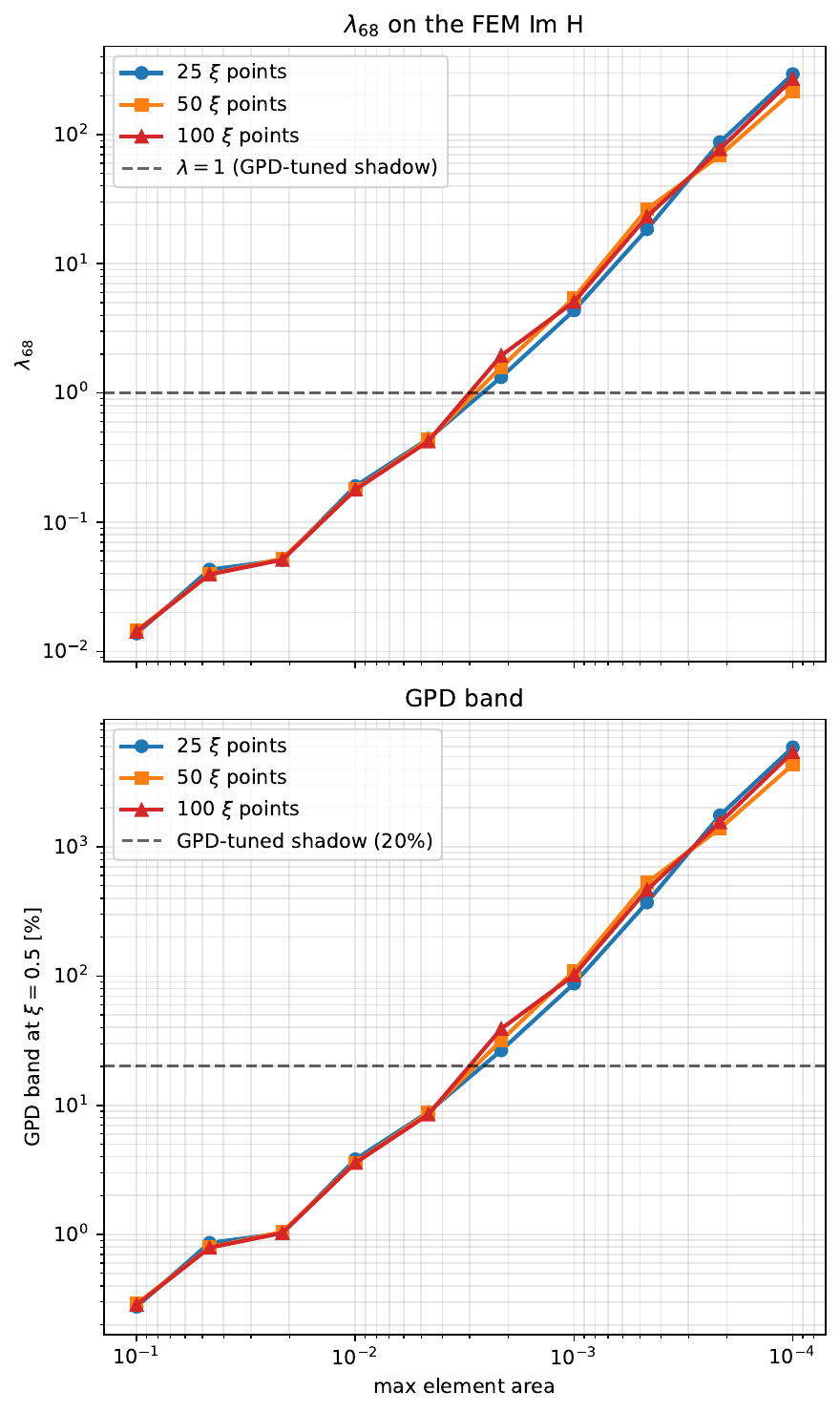}
    \caption{Statistical invisibility of the NLO shadow in $\mathrm{Im}\,\mathcal{H}$ at $Q^2 = 4$~GeV$^2$, as a
function of the maximal element area, for 25, 50 and 100 values of $\xi \in [0.05, 0.5]$.
Top: largest $\lambda_{68}$ for which the FEM-reconstructed shadow contributes less than 20\% of the GK
$\mathrm{Im}\,\mathcal{H}$ at $68\%$ of the $\xi$ values.
Bottom: corresponding GPD band at $\xi = 0.5$, computed with the exact shadow GPD.
Dashed lines: the normalised shadow ($\lambda = 1$, 20\% band).}
    \label{fig:Uncertaintywidth_68}
\end{figure}

\begin{figure}
    \centering
    \includegraphics[width=0.9\linewidth]{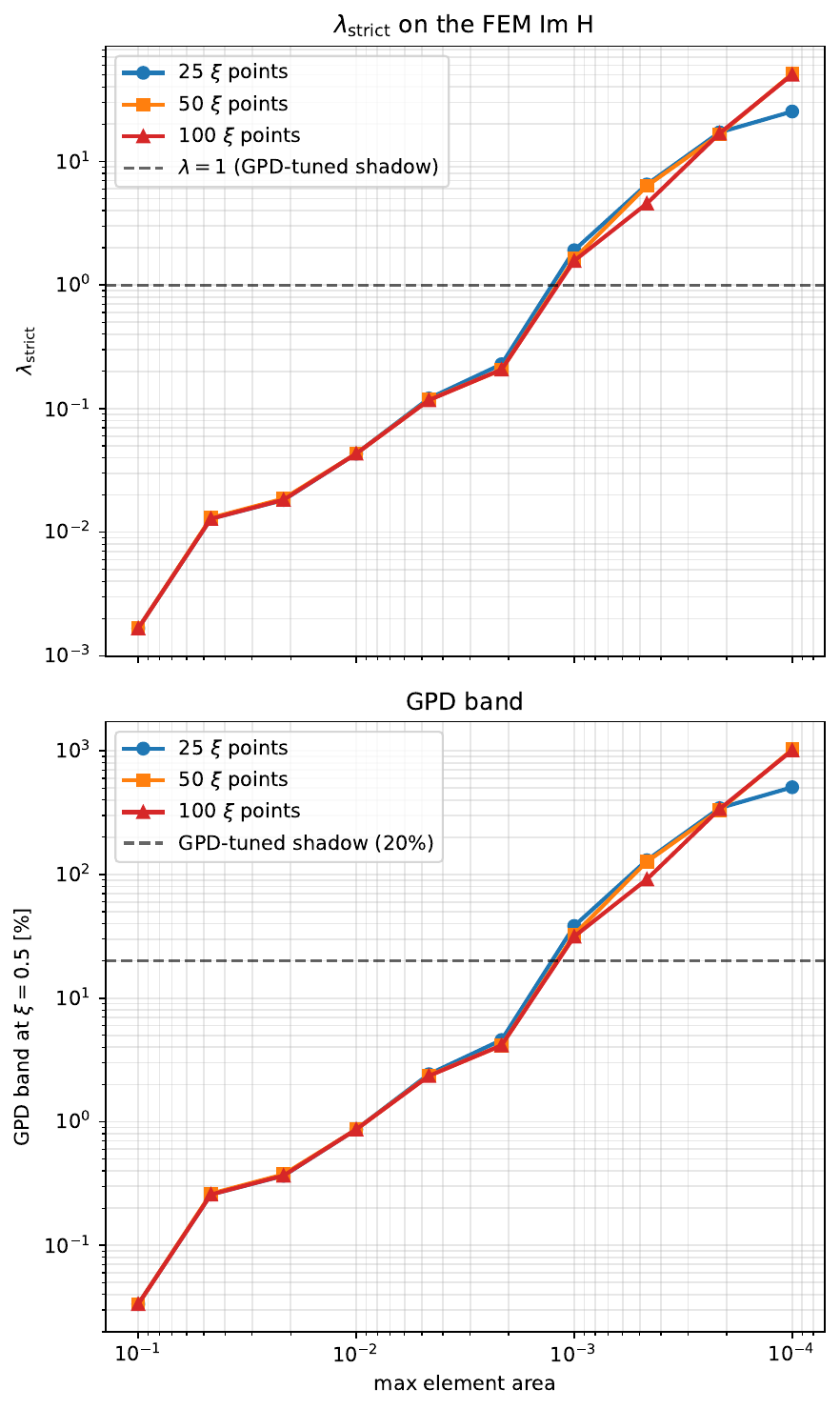}
    \caption{Same as Fig.~\ref{fig:Uncertaintywidth_68} for strict invisibility: the shadow contributes less than 20\% of the GK
$\mathrm{Im}\,\mathcal{H}$ at every sampled $\xi$.}
    \label{fig:Uncertaintywidth_strict}
\end{figure}

The results are presented in Figs. \ref{fig:Uncertaintywidth_68} and \ref{fig:Uncertaintywidth_strict} respectively for $\lambda_{68}$ and $\lambda_{\mathrm{strict}}$.
Several comments can be made.
First, the value of $\lambda$, and consequently the uncertainty band around the GK GPD, grows as a power law, and crosses the 20\% uncertainty range for $a$ around $3\times10^{-3}$ in the statistical case and around $10^{-3}$ in the strict case.
Second, the increase of the density of data points in the range $\xi \in [0.05,0.5]$ has a small impact on the general trend at fixed $Q^2$.
This can be understood as we expect that the larger $\xi$ is, the more constraining Eq. \eqref{eq:LambdaCondition} becomes.
Consequently, once the condition for $\xi=0.5$ is implemented, most of the constraint on $\lambda$ is already in. 
\footnote{We highlight that this feature comes from the choice of a polynomial basis, and from the fact that we impose the shadow GPD to vanish in the forward limit. The result might be different using another basis to probe the null space.}

\begin{figure}
    \centering
    \includegraphics[width=0.9\linewidth]{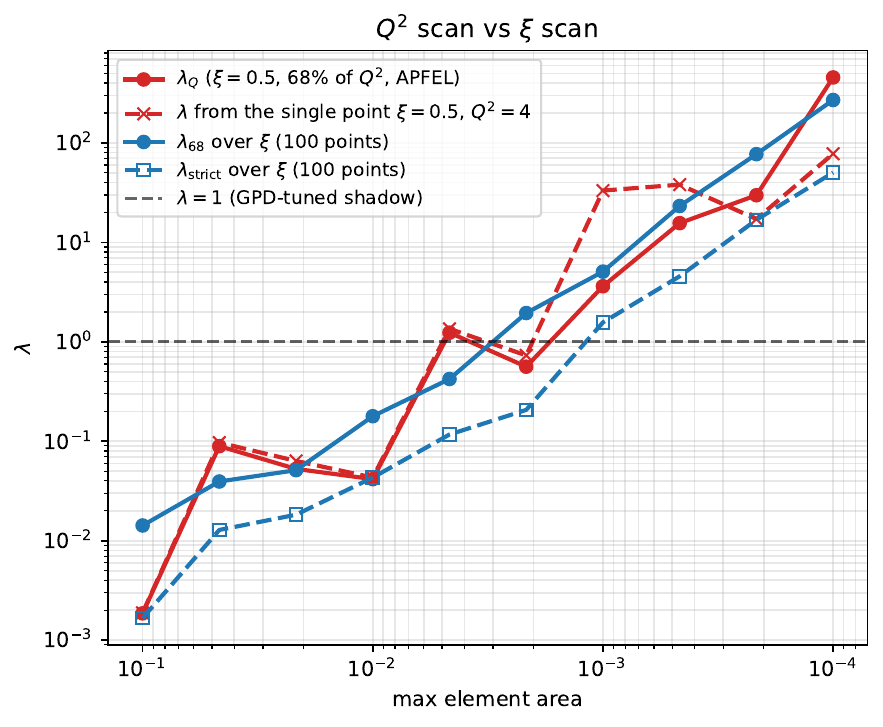}
    \caption{Tuning parameter $\lambda$ versus maximal element area, for the NLO shadow. 
Red: $Q^2$ scan at $\xi = 0.5$ with APFEL evolution \cite{Bertone:2013vaa, Bertone:2017gds,Bertone:2022frx} ($68\%$ of $Q^2 \in [2, 10]$~GeV$^2$, solid) and the single point
$Q^2 = 4$~GeV$^2$ (dashed). Blue: $\lambda_{68}$ (solid) and $\lambda_{\rm strict}$ (dashed) over
$\xi \in [0.05, 0.5]$ at $Q^2 = 4$~GeV$^2$.}
    \label{fig:Uncertaintywidth_Q2}
\end{figure}

This idea is consolidated by working at fixed $\xi$ but varying $Q^2$, as shown in Fig. \ref{fig:Uncertaintywidth_Q2}. It illustrates that fixing $\xi =0.5$ and adding $Q^2$ dependence through LO evolution only marginally reduces the impact of the shadow GPDs.
This is again consistent with our findings from the previous section and confirms the fact that the discretisation error, or resolution $a$, is a much stronger regulator than the splitting functions mixing the $Q^2$ and $x$ dependence together.

Before concluding this section, we emphasise that our results are valid regarding the impact of a single continuum shadow GPD after its discretisation.
We do not make any statement regarding the variance of the discretised reconstruction problem given an experimental uncertainty.
The propagation of experimental uncertainties in the reconstruction problem is the next step to tackle in order to properly control the full GPD reconstruction problem.
This is left for future work.


\section{Conclusion}

In this paper, we have discussed the regularisation properties of the discretisation of DDs with the aim of solving the DVCS inverse problem.
Discretised DDs were previously discussed in the context of the covariant extension, and presented as a solution of the ill-posed incomplete Radon inverse problem \cite{Chouika:2017dhe,Chouika:2017rzs,Chavez:2021llq,DallOlio:2024vjv}.
Here, we extend their use to the DVCS inverse problem at LO and NLO.
We numerically check that, with unconstrained kinematical range and data precision, the discretised DVCS inverse problem is invertible, meaning that the discretised DDs do not belong to the null space of the DVCS kernel, provided that the interpolation polynomials are of  low enough degree.
We have checked a couple of systematic biases coming from the forward limit of GPDs.
First, the forward limit by itself, even probed an unconstrained number of times, is incapable of fully constraining the DD on non-pathological geometry, \emph{i.e.} a discretised framework is able to extract genuine off-forward physics even when provided with the ``continuous" PDF.
However, the discretisation introduces one important reconstruction bias, as the convergence of the FEM towards Regge behaviour of the PDF reveals itself to be rather slow.
In a real data analysis, we thus suggest to stick to the Radyushkin-like form of the DD $F$, splitting it into a PDF, taken from an independent analysis and a profile function.
Finally, we have illustrated that the impact of evolution is in fact much smaller than the one of discretisation for Jefferson Lab kinematics (at least in the resolution range explored), and we have drawn the relation for the specific NLO shadow mode chosen between the resolution $a$ in $(\beta,\alpha)$ space and the width our function explores around the GK model for a fixed uncertainty on the CFF data.

This study should be seen as a first step toward the characterisation of the inverse problem uncertainties within an FEM framework.
Indeed, several points remain to be investigated in order to properly assess the uncertainties in a ``real data" study.
Among them, we can mention for instance the kinematic coverage, which is sparse, even though this point can be mitigated through a first step extraction of the CFFs themselves.
Fitting CFFs can be seen as ``data augmentation" in a machine learning terminology.
Another important point has not been treated here. 
The NLO null space (or at least a finite dimensional reduction of it) has not been systematically explored, and instead we have carefully constructed one NLO shadow GPD, with remarkably regular behaviour on a basis of the 29th degree null-vector polynomials.
This allows us to work on a specific, very relevant case, but prevents us from drawing statistically meaningful conclusions regarding the bias vs. variance trade-off.
Determining an optimal value for $a$ would anyway depend on the experimental data considered.
Continuing in that direction, we can highlight that the present study focuses on fluctuations due to shadow distributions around the GK model used to compute CFFs.
No attempt was performed to assess a complete reconstruction for the pseudo-data, including the discussion regarding experimental uncertainty propagation.
We have only focused on the uncertainties coming from the inverse problem itself.

In view of a real reconstruction attempt in the non-singlet sector, we expect next-to-leading power (NLP) corrections \cite{Braun:2014sta,Braun:2025xlp} to play an important role, as the dilogarithm function appears in the NLP kernel, but is absent in the NLO one.
It also creates a direct dependence on the DD derivatives, forcing an additional smoothing at the DD level.
This would again, on paper, improve the deconvolution problem, provided that, in practice, the data reveals some sensitivity to these corrections.
With Ref. \cite{Mezrag:2026wcf} claiming a first detection of a higher twist effect at the level of the amplitude, it might indeed be the case. 
Optimisation of the mesh could also be discussed.
Namely, beyond the valence region, a logarithmic adaptive mesh in $\beta$ will probably be necessary in order to exploit the data coming from global fits, especially with the difficulties highlighted before regarding the blind reconstruction of the Regge behaviour.
Finally, we believe that, given the significant uncertainties coming from the experimental data, the regulatory power of the FEM alone will most probably not be sufficient in order to stabilise the set of solutions obtained. 
We expect a secondary level of regularisation to be needed.
Among possible examples, one can think of Gaussian Process Regression (GPR) already used for the extraction of GPDs on the lattice \cite{Dutrieux:2026grg}, thin plate regularisation, elastic maps, etc.
Implementing all these points in PARTONS \cite{Berthou:2015oaw} is currently an ongoing task.

\section*{Acknowledgement}

We are grateful to Hervé Dutrieux, Valerio Bertone, Jérôme Bobin and Pawel Sznajder for fruitful discussions. This research was funded, in part, by l’Agence Nationale de la Recherche (ANR), project ANR-23-CE31-0019, and the Spanish MICINN grant PID2023-151418NB-I00. For the purpose of open access, the authors have applied a CC-BY public copyright licence to any Author Accepted Manuscript (AAM) version arising from this submission.

\appendix


\section{Shadow GPDs}
\label{app:shadow}
For the sake of completeness we recall here the family of shadow DDs built in  \cite{Bertone:2021yyz}, from which we take the lowest order, $N=9$, to perform our LO study:
\begin{align}
&F_N^{q(+)}(\beta, \alpha)=  \beta^{N-8}\left[\alpha^8-\frac{28}{9}     \alpha^6\left(\frac{N^2-3 N+20}{(N+1) N}+\beta^2\right)\right. \nonumber\\
&+\frac{10}{3} \alpha^4\left(\frac{N^2-7 N+40}{(N+1) N}+\frac{2\left(N^2-3 N+44\right)}{3(N+1) N} \beta^2+\beta^4\right) \nonumber \\
& -\frac{4}{3} \alpha^2\left(\frac{N^2-11 N+60}{(N+1) N}-\frac{N-8}{N} \beta^2\right. \nonumber \\
& \hspace{2.5cm}  \left. -\frac{N^2-3 N-28}{(N+1) N} \beta^4+\beta^6\right) \nonumber\\
& \left.+\frac{1}{9}\left(1-\beta^2\right)^2\left(\frac{N^2-15 N+80}{(N+1) N}-\frac{2(N-8)}{N} \beta^2+\beta^4\right)\right] .
\end{align}

\bibliographystyle{unsrt}
\bibliography{Bibliography}

\end{document}